\RequirePackage{fix-cm}
\documentclass[twocolumn,epjc3]{svjour3}  
\smartqed  %
\RequirePackage{graphicx}

\usepackage{amsmath}
\usepackage{amssymb}
\usepackage{epstopdf}
\usepackage{color}
\usepackage{mathptmx}
\DeclareGraphicsExtensions{.eps,.pdf,.png,.gif,.jpg,.mht}

\newcommand{\pt}{$p_\textrm{{T}}$}

\journalname{Eur. Phys. J. C}

\date{\today}
\begin{document}

\title{Measurement of the forward charged particle pseudorapidity density in pp collisions at $\sqrt{s} = 8$ TeV using a displaced interaction point}
\author{
The TOTEM Collaboration:
G.~Antchev\thanksref{15} \and%
P.~Aspell\thanksref{8} \and%
I.~Atanassov\thanksref{8,a} \and%
V.~Avati\thanksref{8} \and%
J.~Baechler\thanksref{8} \and%
V.~Berardi\thanksref{5b,5a} \and%
M.~Berretti\thanksref{7b,8} \and%
E.~Bossini\thanksref{7b,7aSec} \and%
U.~Bottigli\thanksref{7b,7aSec} \and%
M.~Bozzo\thanksref{6b,6a} \and%
E.~Br\"{u}cken\thanksref{3a,3b} \and%
A.~Buzzo\thanksref{6a} \and%
F.~S.~Cafagna\thanksref{5a} \and%
M.~G.~Catanesi\thanksref{5a} \and%
C.~Covault\thanksref{9} \and%
M.~Csan\'{a}d\thanksref{4,e} \and%
T.~Cs\"{o}rg\H{o}\thanksref{4} \and%
M.~Deile\thanksref{8} \and%
M.~Doubek\thanksref{1b} \and%
K.~Eggert\thanksref{9} \and%
V.~Eremin\thanksref{b} \and%
F.~Ferro\thanksref{6a} \and%
A. Fiergolski\thanksref{5a,c2} \and%
F.~Garcia\thanksref{3a} \and%
V.~Georgiev\thanksref{11} \and%
S.~Giani\thanksref{8} \and%
L.~Grzanka\thanksref{10,d} \and%
J.~Hammerbauer\thanksref{11} \and%
J.~Heino\thanksref{3a} \and%
T.~Hilden\thanksref{3a,3b} \and%
A.~Karev\thanksref{8} \and%
J.~Ka\v{s}par\thanksref{1a,8} \and%
J.~Kopal\thanksref{1a,8} \and%
V.~Kundr\'{a}t\thanksref{1a} \and%
S.~Lami\thanksref{7a} \and%
G.~Latino\thanksref{7b,7aSec} \and%
R.~Lauhakangas\thanksref{3a} \and%
T.~Leszko\thanksref{c} \and%
E.~Lippmaa\thanksref{2} \and%
J.~Lippmaa\thanksref{2} \and%
M.~V.~Lokaj\'{i}\v{c}ek\thanksref{1a} \and%
L.~Losurdo\thanksref{7b,7aSec} \and%
M.~Lo~Vetere\thanksref{6b,6a} \and%
F.~Lucas~Rodr\'{i}guez\thanksref{8} \and%
M.~Macr\'{i}\thanksref{6a} \and%
T.~M\"aki\thanksref{3a} \and%
A.~Mercadante\thanksref{5a} \and%
N.~Minafra\thanksref{5b,8} \and%
S.~Minutoli\thanksref{6a} \and%
F.~Nemes\thanksref{4,e} \and%
H.~Niewiadomski\thanksref{8} \and%
E.~Oliveri\thanksref{7b} \and%
F.~Oljemark\thanksref{3a,3b} \and%
R.~Orava\thanksref{3a,3b} \and%
M.~Oriunno\thanksref{f} \and%
K.~\"{O}sterberg\thanksref{3a,3b} \and%
P.~Palazzi\thanksref{7b} \and%
Z.~Peroutka\thanksref{11} \and%
J.~Proch\'{a}zka\thanksref{1a} \and%
M.~Quinto\thanksref{5a,5b} \and%
E.~Radermacher\thanksref{8} \and%
E.~Radicioni\thanksref{5a} \and%
F.~Ravotti\thanksref{8} \and%
E.~Robutti\thanksref{6a} \and%
L.~Ropelewski\thanksref{8} \and%
G.~Ruggiero\thanksref{8} \and%
H.~Saarikko\thanksref{3a,3b} \and%
A.~Scribano\thanksref{7b,7aSec} \and%
J.~Smajek\thanksref{8} \and%
W.~Snoeys\thanksref{8} \and%
J.~Sziklai\thanksref{4} \and%
C.~Taylor\thanksref{9} \and%
N.~Turini\thanksref{7b,7aSec} \and%
V.~Vacek\thanksref{1b} \and%
J.~Welti\thanksref{3a,3b} \and%
J.~Whitmore\thanksref{h} \and%
P.~Wyszkowski\thanksref{10} \and%
K.~Zielinski\thanksref{10}%
}

\thankstext{e1}{Corresponding author's e-mail: mirko.berretti@cern.ch}

\institute{Institute of Physics of ASCR, Praha, Czech Republic,\label{1a} \and
Czech Technical University, Praha, Czech Republic,\label{1b} \and
National Institute of Chemical Physics and Biophysics NICPB, Tallinn, Estonia,\label{2} \and
Helsinki Institute of Physics, Helsinki, Finland,\label{3a} \and
Department of Physics,  University of Helsinki, Helsinki, Finland,\label{3b} \and
MTA Wigner Research Center,  RMKI Budapest, Hungary,\label{4} \and
INFN Sezione di Bari, Bari, Italy,\label{5a} \and
Dipartimento Interateneo di Fisica di  Bari, Italy,\label{5b} \and
INFN Sezione di Genova, Genova, Italy,\label{6a} \and
Universit\`{a} degli Studi di Genova,  Genova, Italy,\label{6b} \and
INFN Sezione di Pisa, Pisa, Italy,\label{7a} \and
Universit\`{a} degli Studi di Siena and Gruppo Collegato INFN di Siena,  Siena, Italy,\label{7b} \and
CERN, Geneva, Switzerland,\label{8} \and
Case Western Reserve University,  Dept. of Physics, Cleveland, OH, USA,\label{9} \and
AGH University of Science and Technology, Krakow, Poland,\label{10} \and
University of West Bohemia, Pilsen, Czech Republic,\label{11} \and
Ioffe Physical - Technical Institute of Russian Academy of Sciences, St.Petersburg, Russia,\label{b} \and
SLAC National Accelerator Laboratory, Stanford CA, USA,\label{f} \and
Penn State University, Dept.~of Physics, University Park, PA USA,\label{h} \and 
INRNE-BAS, Institute for Nuclear Research and Nuclear Energy, Bulgarian Academy of Sciences, Sofia, Bulgaria,\label{15} \and
Warsaw University of Technology, Warsaw, Poland.\label{c} \and
Also at INRNE-BAS, Institute for Nuclear Research and Nuclear Energy, Bulgarian Academy of Sciences, Sofia, Bulgaria,\label{a} \and
Also at INFN Sezione di Pisa, Pisa, Italy,\label{7aSec} \and
Also at Warsaw University of Technology, Warsaw, Poland,\label{c2} \and
Also at Institute of Nuclear Physics, Polish Academy of Science, Cracow, Poland,\label{d} \and
Also at Department of Atomic Physics, E\"otv\"os University,  Budapest, Hungary,\label{e} \and
Also at Penn State University, Dept.~of Physics, University Park, PA USA.\label{g}
}

\maketitle

\abstract{
The pseudorapidity density of charged particles dN$_{\textnormal{ch}}$/d$\eta$
is measured by the TOTEM experiment in pp collisions at $\sqrt{s} =$ 8 TeV
within the range $3.9<\eta<4.7$ and $-6.95<\eta<-6.9$. Data were collected in a
low intensity LHC run with collisions occurring at a distance of 11.25 m from
the nominal interaction point. The data sample is expected to include  96-97\%
of the inelastic proton-proton interactions. The measurement reported here
considers charged particles with $p_T>0$ MeV/c, produced in inelastic
interactions with at least one charged particle in $-7<\eta<-6$ or
$3.7<\eta<4.8$.  The dN$_{\textnormal{ch}}$/d$\eta$ has been found to decrease
with $|\eta|$, from 5.11 $\pm$  0.73 at $\eta=$3.95 to 1.81 $\pm$ 0.56 at
$\eta=-$6.925. Several MC generators  are compared to the data and are found to
be within the systematic uncertainty of the measurement.
\PACS{
      {13.85.Hd }{ Inelastic scattering: many-particle final states}   %
     } %
}

\section{Introduction}
The pseudorapidity density of  charged particles produced in high energy proton-proton (pp) collisions is a key observable for the characterization of the hadronic final state. 
Non-perturbative models are used in Monte Carlo (MC) event generators to describe the soft-QCD dynamics of the hadronic interaction~\cite{Skands:2010ak,Ryskin:2011qe}. In the forward region, where diffractive interactions are important, beam remnant and underlying event activity make  the uncertainty on the particle production even more pronounced. Direct measurements of forward pseudorapidity distributions are therefore valuable in constraining the theoretical models. A better knowledge of these effects is also important for the interpretation of the high energy air showers produced by cosmic rays~\cite{Engel:2002id,Albrow:2006xt,d'Enterria:2011kw}.

This work reports the measurement of the charged particle pseudorapidity
density (dN$_{\textnormal{ch}}$/d$\eta$) at $\sqrt{s} =$ 8 TeV in the ranges
3.9$<\eta<$4.7 and  $-6.95<\eta<-6.9$. The measurement is  obtained for a
sample of events recorded with a minimum bias trigger in pp collisions
displaced by 11.25 m from the nominal interaction point (IP) location. These
events have at least one charged particle with either $3.7<\eta<4.8$ or
$-7<\eta<-6$ and are corrected to include charged particles with transverse
momentum down to $p_T =$ 0 MeV/c. dN$_{\textnormal{ch}}$/d$\eta$ is here
defined as the mean number of charged particles per single pp collision and
unit of pseudorapidity $\eta$, where
$\eta\,\equiv\,-\textnormal{ln}[\textnormal{tan}(\theta/2)]$, and $\theta$ is
the polar angle of the direction of the particle with respect to the
anticlockwise beam direction.  The analysis reported here  follows closely the
ones reported in ~\cite{Aspell:2012ux,Chatrchyan:2014qka}.

\section{Experimental apparatus and track reconstruction}
\label{sec:Event_selection}

The TOTEM experiment~\cite{Anelli:2008zza,Antchev:2013hya} is composed of three subdetectors: the Roman Pot  detectors and the T1 and  T2 telescopes. The related right-handed coordinate system has the origin at the nominal interaction point 5 (IP5) of LHC, the $x$-axis pointing towards the centre of the accelerator, the $y$-axis pointing upwards, and the $z$-axis pointing along the anticlockwise-beam direction. The azimuthal angle, $\phi$, is measured in the $(x,y)$ plane, where $\phi=0$ is the  $+x$ and $\phi=\pi/2$ is the $+y$ direction. Inelastic events are triggered by the two T2 telescopes, which are placed symmetrically on both sides of the nominal IP5 at about $|z|=14$~m. Hereafter the T2 telescope covering the positive (negative) pseudorapidities will be referred as T2$+$ (T2$-$). Assuming standard collisions at the nominal IP5, they detect charged particles produced in the pseudorapidity range $5.3<|\eta|<6.5$, with full azimuthal acceptance.  One telescope consists of two half-arms, with each half-arm composed of 10 semicircular planes of triple-GEM (gas electron multiplier) chambers~\cite{Bagliesi:2010zz}, arranged within 40~cm length space along the $z$-axis. 
Each chamber provides two-dimensional information on the track position, covering 192$^{\circ}$ of azimuth angle  with a small overlap region along the vertical axis between chambers of two neighboring half-arms. 
Every chamber has a double-layered read out board containing two columns of 256 concentric strips (400~$\mu$m pitch, 80~$\mu$m width) to measure the radial coordinate and a matrix of 1560 pads, each covering 
$\Delta\eta\times\Delta\phi\approx$~0.06$\times$0.018~rad, to measure the azimuthal coordinate and for triggering. 
The radial and azimuthal coordinate resolutions are about 110~$\mu$m and 1$^{\circ}$, respectively. 
The detailed MC simulations of the TOTEM detectors are based on \textsc{Geant}4~\cite{Agostinelli:2002hh}. Simulated events are processed and reconstructed in the same manner as collision data. The MC corrections are obtained with  the \textsc{Pythia}8 (tune 4C)~\cite{Sjostrand:2007gs,Corke:2010yf} and \textsc{Sibyll 2.1}~\cite{Ahn:2009wx} generators, hereafter referred as \textsc{Pythia}8 and \textsc{Sibyll}.

The T2 track reconstruction is based on a Kalman filter-like algorithm, simplified thanks to the small amount of material in the GEM planes and the weak magnetic field in the T2 region. The particle trajectory can, therefore, be successfully reconstructed with a straight-line fit. Dedicated algorithms were developed in order to correct for effects due to misalignment of the T2 detector. 
The $x$ and $y$ shifts of the T2 half-arms with respect to the nominal positions and their tilts in the $xz$ and $yz$ planes are determined with a precision respectively of $\sim$1 mm and of 0.3-0.4 mrad.
More details on the tracking algorithm and on the alignment procedures can be found in \cite{Berretti:2012efa}.

The analysis reported in this work is obtained with collisions occurring at 11.25 m from the nominal IP5. The events are therefore asymmetric with respect to T2, whose acceptance is expected to be $3.7<\eta<4.8$ and $-7<\eta<-6$, for T2$+$ and T2$-$, respectively. Events with charged particles produced in this range are expected to be triggered with high efficiency by T2 (see section \ref{dataAN} for more details). However, only particles with $3.9<\eta<4.7$ and $-6.95<\eta<-6.9$ cross a minimal amount of material and are safely distant from any detector borders. These tracks are therefore expected to be efficiently reconstructed and can be recognized to come from the interaction region. 
Simulation studies based on  \textsc{Pythia}8  showed that single tracks are reconstructed with an efficiency $>90$\% for \pt $\,>20$ MeV/c in both the T2$+$ and T2$-$ measurement range. The fraction of primary particles with \pt $\,<20$ MeV/c generated in the acceptance of T2$+$ or T2$-$ is below 1\%. %
The $\eta$-resolution in the measured T2$+$ (T2$-$) range is better than 0.05 (0.03), once the track is identified as coming from the interaction region (see section \ref{sec:Primarysel}).
The pseudorapidity of a track in T2 is defined as the average pseudorapidity of all T2 track hits, calculated from the angle between the $z$-axis and the line joining the hit and the displaced IP.
This definition is adopted on the basis of MC simulation studies and gives an optimal estimation of the pseudorapidity of the particles produced at the IP. 

\section{Data sample}\label{Data Sample}
The data sample consists of 400k events collected in July 2012 during a run with a non-standard $\beta$* = 90~m optics configuration and with a bunch pair colliding at 11.25 m from the nominal IP5.
The  probability of overlapping pp interactions in the same bunch crossing (pileup) is found to be $\sim$2-3\%, estimated from the trigger rate for the colliding bunch pair. 
The rate of beam gas interactions is expected to be less than 0.5\%.
The minimum bias trigger provided by the TOTEM T2 telescopes, whose  efficiency is discussed in section \ref{secdataANTrigg}, required at least one track candidate (trigger track) in either T2$+$ or T2$-$\cite{Antchev:2013haa}. With this selection, the fraction of inelastic cross section seen by T2 is estimated to be 96--97\% of the total pp inelastic cross section at $\sqrt{s}=8$~TeV, according to \textsc{Pythia}8 and \textsc{Sibyll} generators. These values are  $\sim$2\%  larger with respect to the \textsc{Pythia}8 prediction obtained for collisions in the nominal IP5, while the fraction of events included in \textsc{Sibyll} does not change significantly. %
Data have at least a track in  both T2$+$ and T2$-$ in 80\% of the triggered events. Events having tracks only in T2$-$ (T2$+$) are 9.5\% (10.5\%) of the total sample. These fractions are compatible with \textsc{Pythia}8 predictions within 1\%. \textsc{Sibyll} instead predicts 86\%, 6.5\% and 7.5\% probability for a  triggered event to  have tracks in both T2$+$ and T2$-$, only in T2$-$ and only in T2$+$, respectively. %

\section{Analysis procedure}\label{dataAN}
The pseudorapidity density measurement presented here refers to ``stable'' primary charged particles with a lifetime longer than $3\,\times\,10^{-11}$ s, either directly produced in pp collisions or from decays of particles with shorter lifetimes. Such a definition, consistent with that of previous studies~\cite{Khachatryan:2010xs,Khachatryan:2010us,Aad:2010ac,Aamodt:2010pp,Aamodt:2010pp,Aspell:2012ux,Chatrchyan:2014qka}, considers the decay products of K$^0_S$~ and $\Lambda$~hadrons and all of the charged particles generated by interactions with the material in front and around the detectors as secondary particles. 
The measurement is corrected to take into account all primary charged particles with \pt $\,>0$ MeV/c, see also discussion at the end of Section \ref{sec:Event_selection}.

\subsection{Trigger efficiency}\label{secdataANTrigg}
The effect of the trigger inefficiency on the measurement is firstly determined  by using a MC simulation. The inefficiency of the trigger is mainly due to non-operating and to noisy channels which were not
used for the trigger generation. The list of these non working channels is introduced in the trigger simulation, giving an effect on the 
$dN_{\text{ch}}/d\eta$ 
measurement of only about 0.5\% with respect to a fully efficient trigger. To be sure that the trigger performance is not biased by the asymmetric arrival time of the particles in the T2$+$ and T2$-$, another run which used different time latencies of the trigger with respect the nominal bunch crossing time is also analyzed.  The trigger rates of the two runs are compatible. This allows us to check that the trigger rates are not affected by the different timing configuration characterizing this run with respect to the case where collisions are provided at the nominal IP ($z=$ 0 m). All the events with at least a reconstructed track are considered in the analysis. The probability that a triggered event has at least a reconstructed track is close to 100\%. According to \textsc{Pythia}8 (\textsc{Sibyll}) the triggered  events have a probability of 68.5\% (70\%)  of having primary charged particles in both the T2 telescopes. The probability to have primary charged particles only in T2$-$ is 9\% (11\%), while the probability to have them only in T2$+$ is 17.5\% (18\%). 

\subsection{Primary track selection}\label{sec:Primarysel}
About 80-85\% of the reconstructed tracks in the analysed $\eta$-range of the T2$-$ and T2$+$ telescope are due to secondary particles, mainly electrons and positrons generated by photon conversions or electromagnetic showers in the material. In T2$+$, conversions are mostly generated in the lower edge of the HF calorimeter of CMS and in the beam pipe at $z \,>$ 13 m. In T2$-$, conversions may happen in the  beam pipe material and in the CMS detectors close to the beam line. It is therefore important to discriminate these secondary particles from the primary charged ones.

\begin{figure}[htb!]
\centering
\includegraphics[width=0.45\textwidth,height=0.3\textheight]{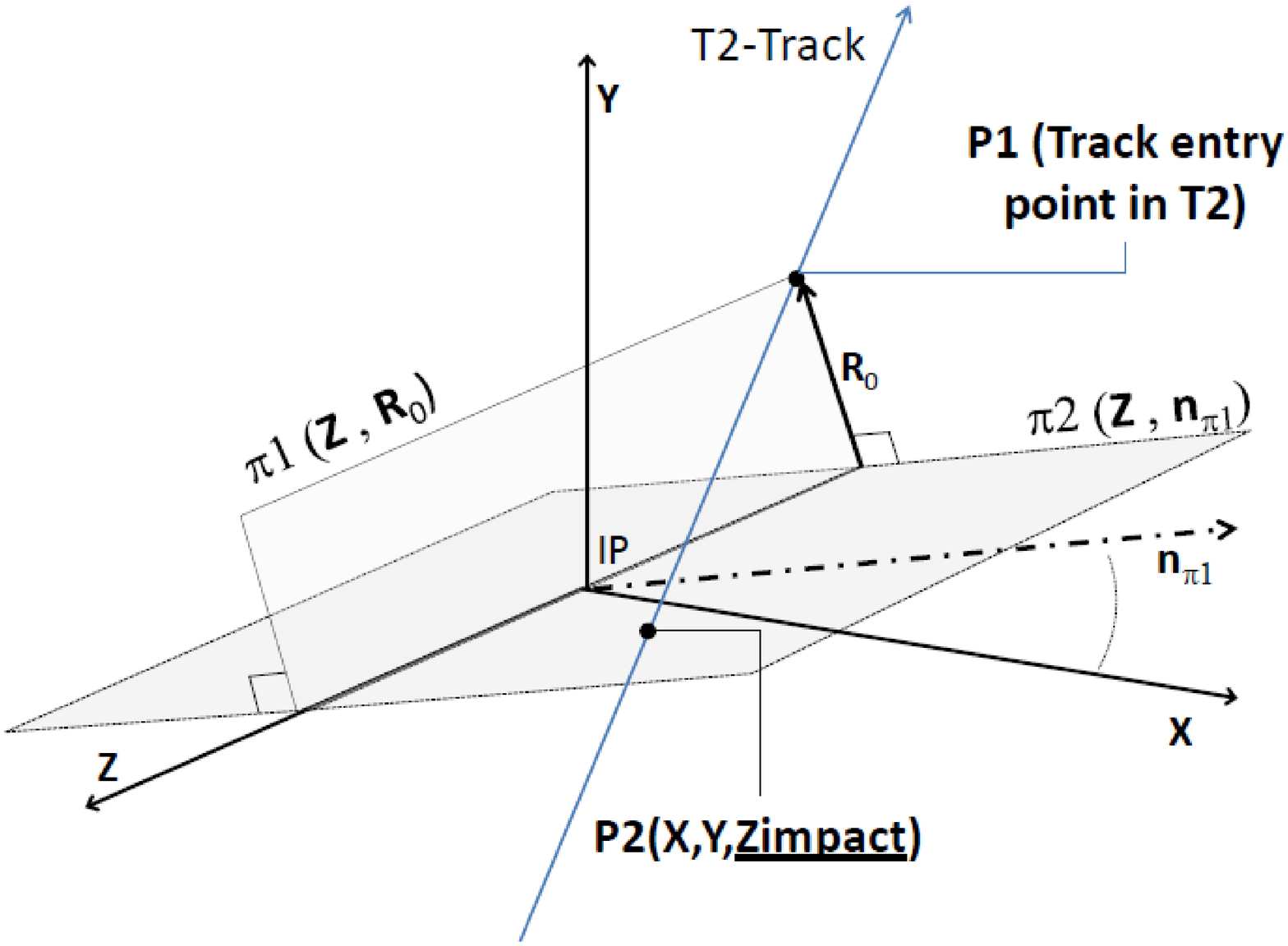}
\caption{Definition of the $z_{\rm impact}$ parameter.}
\label{fig:zimpact}
\end{figure}

In T2$+$, the most effective primary/secondary particle separation is achieved by using the $z_{\rm impact}$ track parameter (see Fig.~\ref{fig:zimpact}), which is defined as the $z$ coordinate of the intersection point between the track and a plane (``$\pi$2'') containing the $z$-axis and orthogonal to the plane defined by the $z$-axis and the
track entry point in T2 (``$\pi$1'')~\cite{Berretti:2012efa}. This parameter is found to be stable against residual misalignment biases.

Simulation studies demonstrated that the $z_{\rm impact}$ distribution can be described by the sum of two Gaussian distributions (hereafter referred to as a ``double-Gaussian'' distribution) mainly due to primary particles, while most of secondary particles with $z_{\rm impact}$ in the primary region can be described by the sum of two exponential distributions (hereafter referred to as a ``double-exponential''). 

\begin{figure}[htb!]
\centering
\includegraphics[width=1.0\linewidth, height=0.3\textheight]{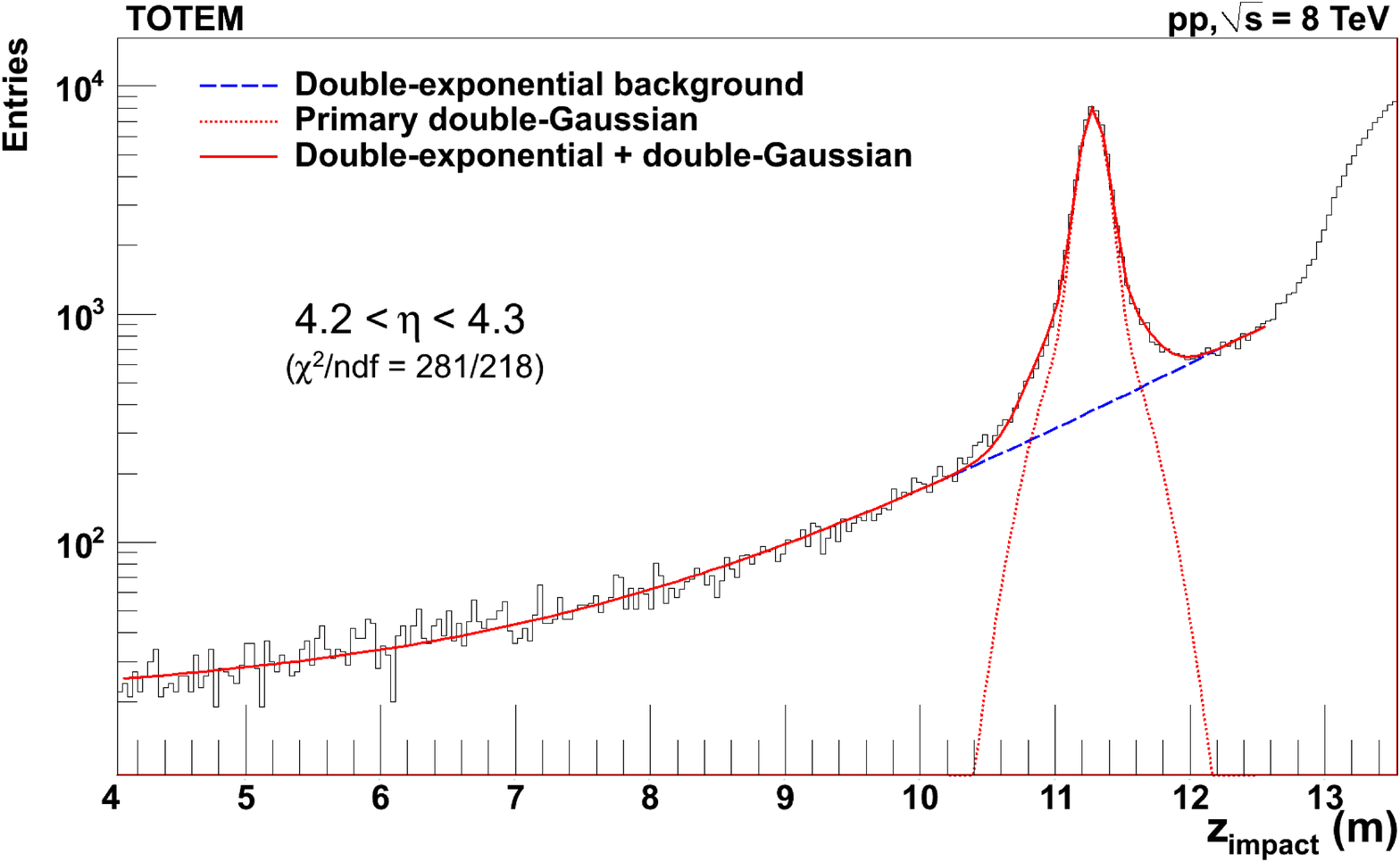}
\caption{The $z_{\rm impact}$ parameter distribution for the data tracks reconstructed in one T2$+$ half-arm in the range $4.2<\eta<4.3$. A global (double-Gaussian + double-exponential function) fit, performed in the range from $4$~m to 12.5~m, is shown by the solid curve. 
The dashed curve represents the double-exponential component from secondary particles, while the dotted curve is the double-Gaussian component, mainly due to primary tracks.
}\label{fig:ZIMPACTDATA}
\end{figure}

Fig.~\ref{fig:ZIMPACTDATA} shows the $z_{\rm impact}$ parameter distribution in one of the central bins of the positive $\eta$ range under study. A combined fit is performed for each $\eta$ bin of the 
$dN_{\text{ch}}/d\eta$ 
distribution with the sum of a double-Gaussian and of a double-exponential function, giving standard deviations (amplitudes) of both Gaussian functions that increase (decrease) with $\eta$.  The mean, required to be the same for both Gaussian distributions, the standard deviations and the amplitudes of the two Gaussian functions as  well as the mean and the amplitude of the exponentials are left free in the fit. The relative abundance  of secondary particles decreases with increasing $\eta$.
Simulations predict a contamination of the double-Gaussian distribution by secondary particles at the level of about 15-20\%. They are mainly given by photons converted in the material between the displaced IP and T2, with a smaller amount of decay products from strange particles. These particles are distributed symmetrically around $z_{\rm impact}=11.25$ m, still following a Gaussian-like distribution.%
The T2$+$ tracks are considered ``primary candidates'' if they satisfy a $z_\text{impact}$ requirement set, for each $\eta$ bin, such that 96\% of the area of the double-Gaussian, symmetric around the mean, is included. %

In order to discriminate primary from secondary tracks in T2$-$ the same strategy as the one described above cannot be used. Indeed, MC studies show that the  $z_\text{impact}$ distribution of the primary particles in T2$-$ is much wider. In this case, a primary to secondary separation based on the $z_\text{impact}$  parameter would  heavily rely on the MC predictions. This worsening on the $z_\text{impact}$ parameter resolution for T2$-$ is due to the  bigger impact that multiple scattering and magnetic field have  on the extrapolation of the track towards the collision region, which is about 25 m away from T2$-$.  Moreover, the impact that the telescope misalignment has on the $z_\text{impact}$ distribution in T2$-$ is expected to be larger as the angles of the primary particles are smaller.%

\begin{figure}[htb!]
\centering
\includegraphics[width=1.0\linewidth,height=0.3\textheight]{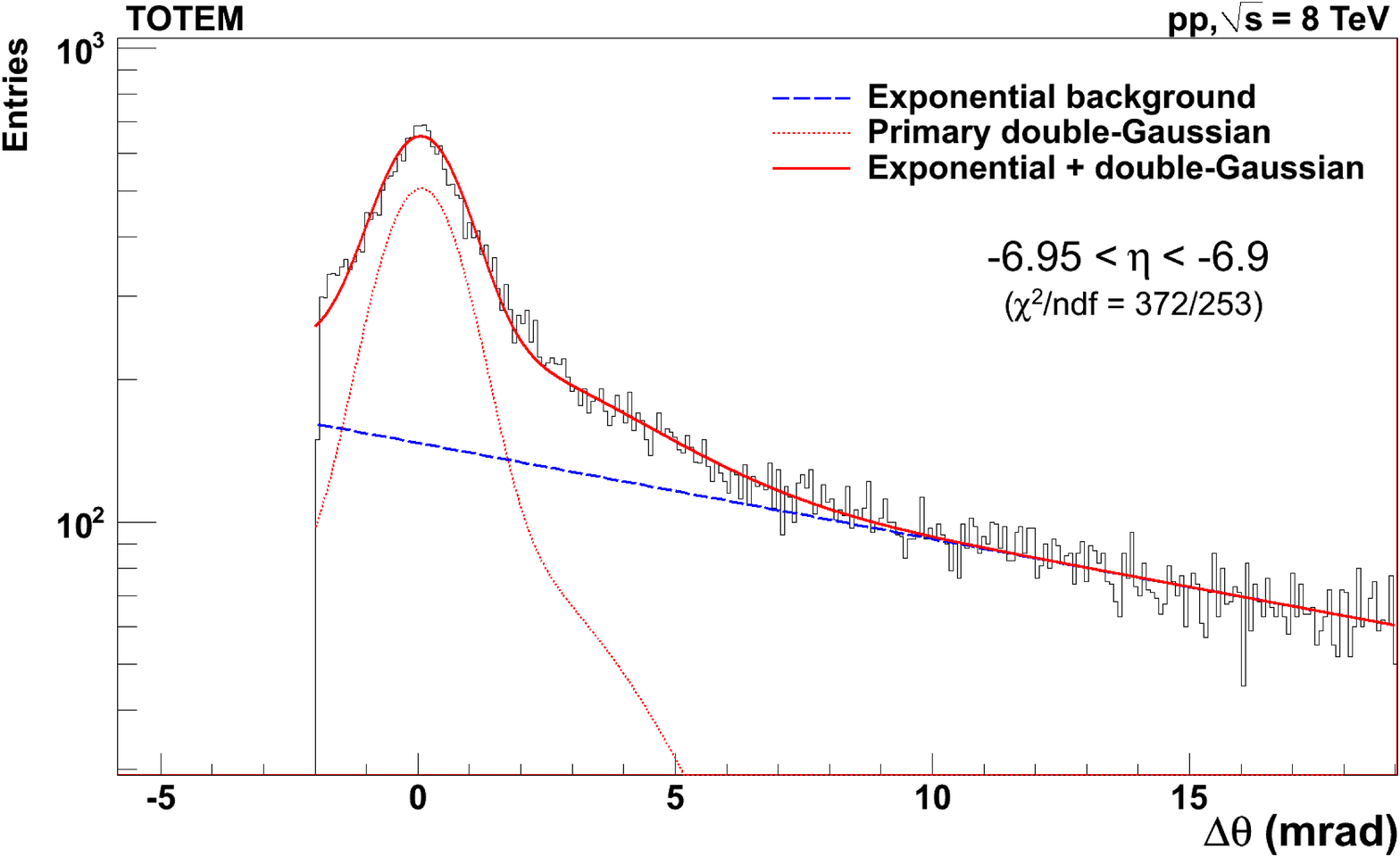}%
\caption{The $\Delta\theta$ parameter distribution for the data tracks reconstructed in  T2$-$. A global (double-Gaussian + exponential function) fit, performed in the range from -2~mrad to 19~mrad, is shown by the solid curve. 
The dashed curve represents the  exponential component from secondary particles, while the dotted curve is the double-Gaussian component, mainly due to primary particles.
}\label{fig:THETAIMPACTDATA}
\end{figure}

A data-driven selection of the primary tracks in T2$-$ is still possible using the $\Delta\theta$ variable. This is  defined as $\Delta\theta = \theta_{fit} - \theta_{IP} $, where $\theta_{IP}$ is the average polar angle of the track deduced from its entry/exit point in the detector (assuming that the particle is coming from the displaced IP) and $\theta_{fit}$ is the absolute value of the polar angle  obtained with a standard fit based on the reconstructed T2 hits.  The choice of this variable is motivated by MC simulation studies.
Fig.~\ref{fig:THETAIMPACTDATA} shows the distribution  of the $\Delta\theta$ parameter obtained  in the $\eta$ region of T2$-$, which is investigated in this work. 

With respect to the $z_\text{impact}$ variable, $\Delta \theta$ has the disadvantage of having  only one side of the distribution that is largely dominated by secondaries. This gives a larger systematic uncertainty related to the subtraction of the secondary contribution. 
However, MC studies show that the peak around $\Delta\theta=0$ mrad is still dominated by primary particles and the full distribution can be fitted by a double-Gaussian function, which mainly contains the primary tracks, and an exponential function which describes the secondaries at large values of $\Delta\theta$. The  parameters of the fit are left unconstrained during the fit procedure. More details about this procedure and on its uncertainty will be reported in sections \ref{sec:T2_dndetaminus} and ~\ref{sec:systUnc}.  According to MC simulations, part of the secondaries doesn't follow the exponential distribution and cannot be separated using the fit of $\Delta\theta$, as they give a symmetric contribution around 0 mrad. The fraction of the double-Gaussian area due to the secondaries is predicted to be about 32\%. %
Similarly to the T2$+$ case, a track in  T2$-$ is considered a ``primary candidate'' if it satisfies a $\Delta\theta$ requirement, set such that 96\% of the area of the double-Gaussian, symmetric around the mean, is included.

\subsection{Event selection correction}
\label{sec:EventSelCorr_dndeta}

In order to take into account the differences between the analysis sample defined at the MC-particle level and the one experimentally selected based on the reconstructed tracks, a correction factor needs to be introduced.
This correction is calculated for each $\eta$ bin from the ratio
\begin{equation}
C_{\text{sel}}(\eta)=\dfrac{\text{d}N_{\text{ch}}/\text{d}\eta_{\text{gen}}|_{\text{gen~selected}}}{\text{d}N_{\text{ch}}/\text{d}\eta_{\text{gen}}|_{\text{reco~selected}}},
\end{equation}
where the numerator is the pseudorapidity density obtained from the MC simulation for events selected based on the charged particles generated within the T2 acceptance at the displaced IP. The denominator is the  density of charged particles arriving in T2, obtained by selecting the simulated events with at least a track reconstructed in T2, as for the data. In general, $C_{\text{sel}}$ is different from unity because of triggered events where only secondary tracks are reconstructed or because of primary charged particles which do not arrive in T2. %
 The $C_{\text{sel}}$  correction factor is evaluated with \textsc{Pythia}8 and \textsc{Sibyll}. Moreover, to quantify possible biases related to this correction, the analysis was repeated requiring that  
 events contain at least a primary candidate track in T2$+$. More details on the numerical values of $C_{\text{sel}}(\eta)$ and on their uncertainties are reported in sections \ref{sec:T2_dndetaplus},
 \ref{sec:T2_dndetaminus} and \ref{sec:systUnc}.

\subsection{Measurement of $\textrm{d}N_{\text{ch}}/\textrm{d}\eta$ in T2$+$}
\label{sec:T2_dndetaplus}

\begin{figure}[htb!]
\centering
\includegraphics[height=3.2in, width=1.0\linewidth]{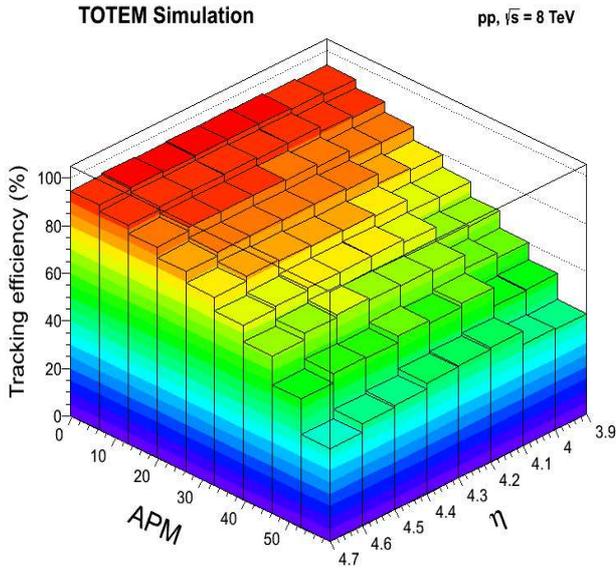}%
\caption{Primary track efficiency as a function of $\eta$ and of the average pad cluster multiplicity per plane
(APM) in one T2$+$ half-arm. The effect of the primary track candidate selection criteria is included in the efficiency.}
\label{fig:T2Effi}
\end{figure}

An analysis similar  to the ones described in  ~\cite{Aspell:2012ux,Chatrchyan:2014qka} has been developed to evaluate the pseudorapidity density in the T2$+$ region. The measurement is performed  for each T2$+$ half-arm independently, thus providing a  consistency check, as each half-arm differs in its alignment and track reconstruction efficiency.  
The number of primary tracks passing the $z_{\rm impact}$ parameter selection criteria is estimated for each $\eta$ bin as a function of the $z_{\rm impact}$ value, using the double-Gaussian and double-exponential fits described in Section~\ref{sec:Primarysel}.
The fraction of primary tracks candidates associated to the double-Gaussian distribution ranges from about 74\% (lower $\eta$ bins) to about 87\% (higher $\eta$ bins), and is used to weight each track by the probability for it to be a primary. Each track is also weighted by the primary track efficiency, which depends on $\eta$ and on the average pad cluster multiplicity per plane (APM) in the corresponding half-arm. The APM probability is a rapidly decreasing distribution, with an average of about 27 and an RMS of about 26. 
The primary track efficiency, evaluated from MC generators, is defined as the probability to successfully reconstruct a generated primary track (with $p_T>0$ MeV/c) that traverses the detector  yielding a $z_{\rm impact}$ parameter within the allowed region. %
Fig.~\ref{fig:T2Effi} shows the primary track efficiency as a function of the track pseudorapidity and of the event APM for one of the T2$+$ half-arms. The  primary track efficiency averaged over APM ranges from about 75\% to about 80\%. 
Additional comparisons of the data and MC track $\chi^2$-probability distributions show that the primary MC efficiencies shown in Fig. \ref{fig:T2Effi} have to be reduced by 2\%.
The rate of multiple associations of reconstructed tracks to the primary one is negligible ($<$0.4\%) once the $z_{\rm impact}$ requirement is imposed.

Conversion of photons from $\pi^{0}$ decays in the material between the displaced IP and T2, as well as decay products of strange particles, also contribute to the double-Gaussian peak. 
The overall non-primary contribution, to be subtracted from the double-Gaussian peak, is estimated as a function of $\eta$ with \textsc{Pythia}8 and \textsc{Sibyll}. %
The value of this correction  ranges from about 17\% (low $\eta$) to 12\% (high $\eta$) and is obtained as the average of the two  MC predictions.  
The correction factor for the event selection bias ($C_{\text{sel}}(\eta)$) is found to be about 1.1 according to \textsc{Pythia}8 and \textsc{Sibyll}. This factor has been obtained after having imposed that both MC reproduce the same relative amount of events with no primary candidates as found in the data. Bin migration effects in $\eta$ are corrected for with \textsc{Pythia}8,  which gives the best description of the slope  of the measured $\textrm{d}N_{\textnormal{ch}}/\textrm{d}\eta$ distribution. The effects are typically at the level of a few percent. %

Events characterised by a high T2 hit multiplicity, typically due to showers generated by particles interacting with the material before T2, are not included in the analysis. These events, where track reconstruction capability is limited, are characterised by an APM value larger than 60 and constituted about 13\% of the sample.  The effect of removing these events is firstly evaluated in a MC study, which resulted in an overall correction factor of about 1.18 (1.28) according to \textsc{Pythia}8 (\textsc{Sibyll}). To verify the stability of this correction an additional method has been developed: the correction is also estimated by extrapolating the measured average multiplicity obtained as function of the maximum APM included in the sample, without  correcting for the excluded fraction of the sample, to APM values above 60. 
The extrapolation, performed with a second degree polynomial, gives a correction of 1.11. The average between the factor predicted from this extrapolation and the one obtained with \textsc{Pythia}8 MC, which better describes the data, is used for this correction. %

The fully corrected $\textrm{d}N_{\text{ch}}/\textrm{d}\eta$ distribution in each $\eta$ bin is determined via: 

\begin{equation}\label{masterformula}
\begin{split}
\vphantom{...}&\frac{\text{d}N_\text{{ch}}}{\text{d} \eta}=\\
& \dfrac{\,\,C_{\text{sel}}(\eta)\sum_{\mathrm{evt, trk \in\, S}}\,\omega_{\mathrm{trk}}(\text{APM},\eta,z_{\rm impact})\,\sum_{j}B_{j}(\eta)}
{\Delta\eta\,N_{\mathrm{evt}}}\,\dfrac{2\pi}{\Delta\phi}
\end{split}
\end{equation}

where S is the sample of tracks with $\eta-\Delta\eta/2 < \eta < \eta + \Delta\eta/2$ satisfying the selection criteria above, $\Delta\eta=0.1$ is the bin width, $C_{\text{sel}}$ is the correction factor related to the event selection (defined in section~\ref{sec:EventSelCorr_dndeta}), $B_j$ is the bin migration correction associated with the $j$-th bin in $\eta$, 
$\Delta\phi/2\pi=192^{\circ}/360^{\circ}$ is the azimuthal acceptance of each T2 half-arm, $N_{\mathrm{evt}}$ is the total number of selected events, and $\omega_{\mathrm{trk}}$ is defined as:
\begin{equation}
\label{t2wtrk}
\omega_{\mathrm{trk}}(\text{APM},\eta,z_{\rm impact})\,=\,
\dfrac{P_\text{prim}(\eta,\mathit{z_{\rm impact}})\,S_{\text{np}}(\eta)\,C_\text{mult}(\eta)}{\epsilon(\eta, \mathrm{APM})},
\end{equation}
where $P_\text{prim}$ is the probability for a track to be primary, $\epsilon$ is the primary track efficiency,  $S_{\text{np}}$ is the correction factor for the non-primary contribution to the double-Gaussian peak, and $C_{\text{mult}}$ is the correction factor accounting for the exclusion of events  with APM values above 60.

The $\textrm{d}N_{\textnormal{ch}}/\textrm{d}\eta$ distribution  obtained refers to charged particles with \pt$>0$ MeV/c. 

\subsection{Measurement of $\textrm{d}N_{\text{ch}}/\textrm{d}\eta$ in T2$-$}
\label{sec:T2_dndetaminus}

The analysis of the pseudorapidity density in  T2$-$ is similar to the one in T2$+$ (eq. \ref{masterformula}). Therefore in this section only  the differences with respect to the analysis performed in T2$+$   are mentioned. For T2$-$, the measurement has been restricted to only one $\eta$ bin ($-6.95<\eta<-6.9$) because only in this range the track reconstruction is efficient and reliable.
The selection of the primary track candidates is based on the $\Delta\theta$ variable described in section~\ref{sec:Primarysel}. The related double-Gaussian and the exponential functions are used to weight each track by the probability for it to be primary ($P_\text{prim}(\eta,\Delta\theta)$). The data and the MC fits are required to produce the same value of the ratio between the exponential and the 
double-Gaussian function at $\Delta\theta=-2$ mrad. This requirement reduces potential data-MC differences in the fit results, which are due to different extrapolated value of the exponential function in the primary region. More details on the systematic uncertainty related to the fit procedures are reported in section \ref{sec:systUnc}. About 35\%  of tracks with $\Delta\theta$ in the primary candidate region are associated to the exponential background.  The non-exponential  background included in the primary double-Gaussian peak region is estimated as an average of the \textsc{Pythia}8 and \textsc{Sibyll} MC generator. The results of the LHCf experiment on the photon 
dN/dE distribution~\cite{Adriani:2011nf} are taken into account by these two MCs. The non-exponential background affecting the primary candidate region corresponds to about~32\% of the selected signal and it is taken into account by the proper correction factor ($S_{\text{np}}(\eta)$).  The primary track efficiency, parametrized as a function of APM ($\epsilon(\eta,\mathrm{APM})$) when including the effect of the primary track candidate selection criteria, is found to be 70\% on average. This efficiency has been corrected by 10\% due to latency issues leading to a data-MC discrepancy.

The correction factor for the event selection bias ($C_{\text{sel}}(\eta)$) is found to be about 1.02 according to \textsc{Pythia}8 and \textsc{Sibyll}. Events having an APM larger than 60 due to the high secondary particle production constitute 16\% of the sample and the associated MC correction factor ($C_\text{mult}(\eta)$) is 1.03. 

To be sure that the analysis results are not biased by the choice of the analysed T2$-$ half-arm and by potential timing issue due to the asymmetric configuration of the run, the measurement is performed by using two different samples. In the run where the latency is optimized for T2$+$, the half-arm in  T2$-$ having the better latency is used. The measurement is then repeated using an ancillary run, where the latency in the T2$-$ is optimal for the other half-arm. As in this case the latency is not optimized for the T2$+$, the 
dN$_{\textnormal{ch}}$/d$\eta$
value has to be corrected for trigger losses due to events with particles only  in T2$+$. This correction is about 10\%. The final result is obtained by averaging the measurements from  the two different runs.

\section{Systematic uncertainties}
\label{sec:systUnc}

The systematic uncertainty evaluation for the $\textrm{d}N_{\textnormal{ch}}/\textrm{d}\eta$ distributions is performed in a similar way as in~\cite{Chatrchyan:2014qka}. In the following details are given  only for the most significant contributions.

In the T2$+$ region, the systematic uncertainty in the $P_\text{prim}$ function, of about 5-6\%, is evaluated by taking into account three effects: a) the sensitivity to the misalignment corrections (2\%), quantified by varying the corrections within their uncertainties, b) the sensitivity to the $z_{\rm impact}$ parameter fitting range (5\%), which was changed by $\pm$0.5 m, and c) the effect of possible deviations of the fitting function for the track $z_{\rm impact}$ distribution (about 2\%). In  T2$-$ the leading contribution to the error of the $P_\text{prim}$ function is given by the fit uncertainty, evaluated by changing the fitting interval used for the exponential fit in the secondary region and without imposing any constraint at $\Delta\theta=-2$ mrad. Since it is difficult to model the background in this region, a conservative approach has been used, where the extreme right point of the fit has been changed from 12 to 22 mrad, resulting in a 20\% fit uncertainty. %

The systematic uncertainty due to non-primary tracks included in the double-Gaussian once the exponential contribution has been removed ($S_{\text{np}}$) is evaluated by taking into account two effects: a) the range of the MC predictions  (about 3\% and 7\% in T2$+$ and T2$-$ respectively), b) the data-MC discrepancy on the ratio between the double-Gaussian and the exponential curve in the  primary candidate region (about 4\% and 7\% in the T2$+$ and T2$-$ respectively). In the T2$-$ these contributions are obtained keeping the relative constraint between the data and the MC fit, as described in \ref{sec:T2_dndetaminus}.

In addition, simulation studies are also performed by varying the thickness of the material in front of T2 by 40\%. This part of the material is the main source of secondary tracks that contribute to the double-Gaussian. The effect of the change of the material results in a possible bias of less than 3\%.
The systematic uncertainty in the primary-track efficiency ($\epsilon$) is evaluated in studies where tracks are reconstructed with a set of five consecutive detector planes (out of the total of ten) in a single T2 half-arm. These tracks are used to determine the track reconstruction efficiency of the other set of detector planes in the same half-arm.
The difference between the simulation and data results obtained with the  above method, is found to be about 5\% for  T2$+$ and about 20\% for  T2$-$ and taken as estimate of the systematic uncertainty.  For T2$-$ the uncertainty is larger due to residual latency issues.

For the T2$+$ analysis, the uncertainty in the correction for the exclusion of events with high secondary-particle multiplicity  ($C_\text{mult}$) is estimated by taking into account the difference between the \textsc{Sibyll} and \textsc{Pythia}8 estimates, and the result of the data-driven extrapolation procedure. The associated uncertainty, about 8\%, is taken as half of the maximum difference among the three predictions. In the T2$-$ region, high multiplicity events are less rich in primary particles and the correction for the excluded events is smaller. The difference between the MC predictions, taken as uncertainty, is about 2\%. 

The uncertainty on the correction accounting for the event selection ($C_{\text{sel}}$)  is evaluated by taking into account both the difference between the corrections from the two MC generators mentioned above and the dependence of the $\textrm{d}N_{\text{ch}}/\textrm{d}\eta$ from the event selection criteria as described in section \ref{sec:EventSelCorr_dndeta}. The overall systematic uncertainty  is found to be less than 3\%.

The maximum discrepancy between the results obtained in each half-arm, taken as additional systematic uncertainty, is found to be 4\% in the T2$+$ and  8\% in the T2$-$. The statistical uncertainty is less than 1\%. Table~\ref{onebinsysttab} shows the statistical and the main systematic uncertainties of the measurement. 
The total uncertainty is obtained by adding in quadrature the reported systematic errors  with the statistical one. A final uncertainty of 13-14\% (31\%) is obtained for the measurement in T2$+$ (T2$-$).

\begin{table*}
\begin{center}
\caption{Systematic and statistical uncertainties in the $\textrm{d}N_{\textnormal{ch}}/\textrm{d}\eta$ measurements for the regions $3.9<\eta<4.7$ and  $-6.95<\eta<-6.9$. 
}
\label{onebinsysttab}
\begin{tabular}{l c c }
\multicolumn{2}{c}{} \\
\hline    
  \multicolumn{1}{c}{ } & $3.9<\eta<4.7$ & $-6.95<\eta<-6.9$    \\
  \hline
Tracking efficiency data-MC discrepancy               &  5-6\%  &   20\%  \\ 
Primary track selection                                     &  5\%  &   20\%     \\ 
Secondaries in the double-Gaussian peak               &  5\%  &   10\%      \\ 
High-multiplicity events                              &  8\%  &   2\%    \\
Quarter discrepancy                                   &  4\%  &   8\%      \\
Material uncertainty                                  &  3\%  &   3\%      \\
Event selection                                       &  $<$3\%  &   $<$3\%      \\
Statistical uncertainty                               &  $<$1\%  &   $<$1\%      \\
\hline
Total (after averaging half-arms \\
and including minor contributions)               &   13-14\%  &  31\%   \\
\hline
\end{tabular}
\end{center}
\end{table*} 

Additional studies are performed for T2$+$ to further characterize the systematic uncertainties. The  uncertainty on the tracking efficiency and on the primary track selection contribute to the $\eta$-uncorrelated part of the uncertainty, which is between 1 and 6\%. The effect of a possible bias introduced by the systematic uncertainties on the measured values at the beginning  and at the end of the T2$+$ $\eta$ range is estimated to be at most 10\%. The measurement in T2$-$ is largely independent from the one in T2$+$.
For the measurement in the T2$-$, an $\eta$-uncertainty of $\sigma_\eta=0.05$  is assumed, by taking into account both the $\eta$-resolution  and the possible effects that residual misalignments can have on the pseudorapidity estimation.

\section{Results}

\begin{figure*}[htb!]
  \begin{center}
\includegraphics[width=0.9\textwidth, height=0.44\textheight]{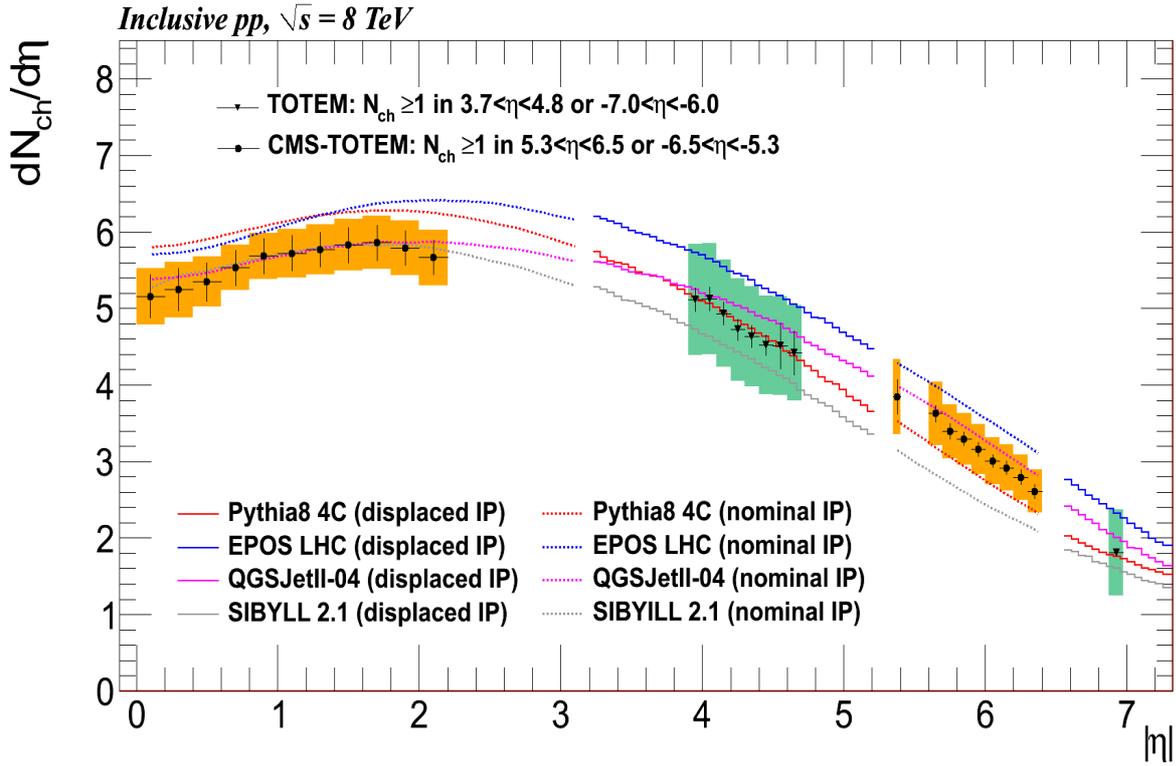}%
\caption{
Charged particle pseudorapidity distributions obtained in pp collisions at $\sqrt{s}=8$ TeV for inelastic events. The coloured bands show the combined systematic and statistical uncertainties and the error bars represent the $\eta$-uncorrelated uncertainties. The results obtained in this work based on collisions at $z=11.25$ m (“displaced IP”) are shown under the green band, while the distributions under the orange band are taken from~\cite{Chatrchyan:2014qka}, where collisions occurred at $z=0$ m (“nominal IP”). The measurements are compared in each $\eta$ region to the corresponding prediction from \textsc{Pythia}8 (tune 4C), \textsc{Sibyll 2.1}, \textsc{Epos} (tune LHC), and {\sc QGSJet}II-04.}
 \label{fig:datacorr}
 \end{center}
\end{figure*}

The  charged particle pseudorapidity distribution measured in this work is presented in Fig.~\ref{fig:datacorr}, together with the results obtained jointly by the CMS and TOTEM Collaborations \cite{Chatrchyan:2014qka} for inelastic events selected in pp collisions at the nominal IP for $\sqrt{s}=$8 TeV.

The green band represents the total uncertainty, while the black error bars are the $\eta$ uncorrelated uncertainties. 
The measurement and  the corresponding MC predictions are shown in bins of $|\eta|$ for a better visualization.
The 
 $\textrm{d}N_{\text{ch}}/\textrm{d}\eta$ 
measured in this work is found to be 5.11 $\pm$ 0.73 at $\eta=3.95$, 4.42   $\pm$ 0.63 at $\eta=4.65$ and 1.81  $\pm$ 0.56 at $\eta=-6.925$,  with negligible statistical uncertainty. The predictions from {\sc QGSJet}II-04~\cite{Ostapchenko:2010vb}, \textsc{Sibyll 2.1}, \textsc{Epos} (tune LHC) ~\cite{Werner:2005jf,Pierog:2013ria}, and \textsc{Pythia}8 (tune 4C) are compatible with the data, even if the \textsc{Sibyll} (\textsc{Epos}) predictions underestimate (overestimate) systematically the data by about 6-10\% (15-30\%). %

The $\textrm{d}N_{\textnormal{ch}}/\textrm{d}\eta$  measured in this work is also reported in table \ref{tabdndeta}, with the corresponding total and $\eta$-uncorrelated uncertainty.

\begin{table}
\caption{The TOTEM $\textrm{d}N_{\textnormal{ch}}/\textrm{d}\eta$ measurement for inelastic pp events with displaced interaction point at $\sqrt{s}=8$ TeV. The reported values represent the
average of two half-arms with the corresponding full systematic (syst) and $\eta$-uncorrelated systematic ($\eta$-uncorr syst) error. The statistical error is negligible. $\eta_0$ represents the central pseudorapidity value in each eta bin. The bin width is 0.05. No value for $\eta$-uncorr syst is quoted for the $\eta<0$ measurement, since it is largely independent from the $\eta>0$ measurements.}
\label{tabdndeta}
\begin{center}
\begin{tabular}{cccc}
   \hline
$\eta_0$ $\,\,$ & 
 $\textrm{d}N_{\text{ch}}/\textrm{d}\eta$ 
& syst error & $\eta$-uncorr syst error\\
    \hline
-6.925  &  1.81  &  0.56  &   -  \\
 3.95 &  5.11  &  0.73  &   0.15  \\
 4.05  & 5.13  &  0.73  &   0.15  \\
 4.15  & 4.93   &  0.70  &  0.15   \\
 4.25  & 4.72  &  0.67  &   0.14  \\
 4.35  & 4.64  &  0.66  &    0.14 \\
 4.45  &  4.52  &  0.64  &  0.14   \\
 4.55  &  4.51  &  0.64  &  0.29   \\
 4.65  &  4.42  &  0.63  &   0.29  \\
\end{tabular}
\end{center}
\end{table}

\section{Summary}
In this work, the measurement of the charged particle pseudorapidity densities
in the ranges 3.9$<\eta<$4.7 and  $-6.95<\eta<-6.9$, for proton-proton
collisions at a centre-of-mass energy of 8 TeV has been reported. The data were
collected using the minimum bias trigger of the TOTEM T2 detector, during a
dedicated run at low intensity and with a non-standard $\beta$* = 90~m optics
configuration. Collisions were provided at a distance of 11.25 m from the
nominal interaction point, allowing  T2 to cover a pseudorapidity range which
is very different from its nominal one.  The measurement has been made
considering charged particles with $p_T>0$ MeV/c, in an inelastic sample with
at least one charged particle produced in either $-7<\eta<-6$ or
$3.7<\eta<4.8$.  Predictions obtained with different MC event generators and
tunes have been found to be consistent with the measurement.

\section*{Acknowledgements}
We thank Benedetto Gorini, Emilio Meschi and the LHC machine coordinators for providing us the dedicated TOTEM runs and for keeping the collisions at $z = 11.25$ m that allowed us to make this measurement.
We are very grateful to the CMS collaboration for providing us the software framework where all the toolkits used for the analysis reported here have been developed.
This work was supported by the institutions listed on the front page and partially also by NSF (US), the Magnus Ehrnrooth foundation (Finland), the Waldemar von Frenckell foundation (Finland), the Academy of Finland, the Finnish Academy of Science and Letters (The Vilho, Yrj\"{o} and Kalle V\"{a}is\"{a}l\"{a} Fund), the OTKA grant NK 101438 (Hungary) and the Ch. Simonyi Fund (Hungary).

\bibliography{CMS_TOTEM_dN_dEta_ARC}{}%
\bibliographystyle{unsrt}

\end{document}